 \documentclass[aip,pop,reprint]{revtex4-1}
 \usepackage{dcolumn}
 \usepackage{bm}
 \usepackage{amsmath}
 \usepackage{tabularx}

 \usepackage{graphicx}
 \usepackage{color}
 \usepackage{epsfig}
 \usepackage{float}
 \usepackage{ulem}

\newcommand{\pp}[2]{\frac{\partial #1}{\partial #2}}
\renewcommand{\vec}[1]{\boldsymbol{#1}}

\newcommand{\y}{\textsc{y}}

\newcommand{\kT}{k_{\textsc b}T}
\newcommand{\scr}{\mathrm{scr}}

\newcommand{\msqrt}{\textsc{sqrt}}
\newcommand{\mmass}{\textsc{mass}}
\newcommand{\msvt}{\textsc{svt}}
\newcommand{\tallfrac}[2]{\frac{\displaystyle #1}{\displaystyle #2}}

\begin{document}


\title{Pair Correlation Functions of Strongly Coupled Two-Temperature Plasma}


\author{Nathaniel R.~Shaffer, Sanat Kumar Tiwari, Scott D.~Baalrud}

\affiliation{Department of Physics and Astronomy, University of Iowa, Iowa City, IA 52242}


\date{\today}

\begin{abstract}

  Using molecular dynamics simulations, we perform the first direct tests of three proposed models for the pair correlation functions of strongly coupled plasmas with species of unequal temperature.
  The models are all extensions of the Ornstein-Zernike/hypernetted-chain theory used to good success for equilibrium plasmas.
  Each theory is evaluated at several coupling strengths, temperature ratios, and mass ratios for a model plasma in which the electrons are positively charged.
  We show that the model proposed by Seuferling, Vogel, and Teopffer~[Phys.~Rev.~A \textbf{40}, 323 (1989)] agrees well with molecular dynamics over a wide range of mass and temperature ratios, as well as over a range of coupling strength similar to that of the equilibrium HNC theory.
  The SVT model also correctly predicts the strength of interspecies correlations and exhibits physically reasonable long-wavelength limits of the static structure factors.
  Comparisons of the SVT model with the Yukawa OCP model are used to show that ion-ion pair correlations are well described by the YOCP model up to $\Gamma_e \approx 1$, beyond which it rapidly breaks down.

\end{abstract}



\maketitle


\section{Introduction}
\label{sec:intro}

Strongly coupled plasmas produced in experiments are often far from thermal equilibrium.
Ultracold neutral plasmas and the dense plasmas in sonoluminescent bubbles, for instance, can have electron and ion temperatures that differ by an order of magnitude or more~\cite{KillianPRL1999,PuttermanARFM2000}.
In inertial confinement fusion plasmas and ultracold plasma mixtures, there can be significant differences in temperatures not just between ions and electrons, but also between the different species of ions~\cite{BergesonDPP2016,RinderknechtPRL2015}.
The combination of strong coupling and multiple temperatures makes these plasmas especially challenging to model, since one cannot freely call upon results from equilibrium statistical mechanics to make predictions about the plasma's thermodynamic and transport properties.
One approach for strongly coupled, two-temperature plasmas is to extend integral equation theories for the equilibrium pair distribution functions to allow multiple temperatures.
In this work, we present the first direct comparisons of three such extensions and evaluate their accuracy against molecular dynamics (MD) simulations.

The pair distribution functions are normally applied in the context of thermal equilibrium, where they can be used to evaluate the pressure, internal energy, and other thermodynamic state variables using exact formulas from equilibrium statistical mechanics.
In non-equilibrium plasmas -- especially those far from equilibrium -- they are used in extensions of ideal gas kinetic theory to treat strongly coupled plasmas.
The pair distributions enter into these models in the form of effective scattering potentials~\cite{BaalrudPRL2013} or local field corrections~\cite{IchimaruPRA1985,DharmaWardanaPRE1998,DaligaultPRE2009,VorbergerPRE2010,BenedictPRE2017}, designed to take approximate account of how the collisional transfer of momentum and energy in the plasma is affected by the many-body physics of strong coupling.
The pair distribution functions of two-temperature systems are also useful for testing the range of validity of approximate one-component models of strongly coupled plasmas.
Most importantly, by treating electrons and ions on equal footing, two-component plasma models grant access to electron-ion transport physics that lie beyond the scope of a one-component treatment.

The present work makes use of a model plasma consisting of ions and positively charged electrons.
This approach is useful in both modeling and simulation (e.g., Ref.~\onlinecite{DaligaultPRE2009}) to circumvent the collapse (recombination) of classical electron-ion plasmas, which to date must be treated with softened electron-ion pseudopotentials.
By instead using positively charged electrons, we are able to isolate the relevant two-temperature physics, which should not depend on the sign of the charge.
Future work will use a recently developed method for modeling strongly coupled electron-ion plasmas~\cite{TiwariPRE2017} to explore the effect of negatively charged electrons on pair correlations and transport.
Notwithstanding, the results shown here are immediately applicable to ionic mixtures with unequal temperatures.

At weak coupling, the pair distribution functions are accurately described by the Debye-H\"uckel theory of electrolytes~\cite{DebyePZ1923}.
For strongly coupled plasmas, however, the triplet and higher-order correlations ignored in the Debye-H\"uckel approach become important.
At thermal equilibrium, these correlations are well approximated by integral equation methods developed from equilibrium statistical mechanics.
The most successful of these involve solving the Ornstein-Zernike (OZ) relations together with an approximate closure, e.g., the hypernetted-chain (HNC) approximation~\cite{HansenMacDonald}.
When the plasma has more than one temperature, two methodologies have been explored: (a) to map the multi-temperature plasma to an effective one-temperature plasma or (b) to extend equilibrium integral equation theories to allow multiple temperatures.

The canonical example of the mapping approach is the Yukawa one-component plasma (YOCP) model.
In the YOCP model, the plasma is partitioned into a strongly coupled component of interest and a weakly coupled background.
All the physics of this background are condensed into a constant screening parameter that modifies the interaction between the strongly coupled particles.
These conditions are realized in dusty plasmas and in present-day ultracold neutral plasma experiments, where the YOCP model has been successfully applied to study the dust and ions, respectively~\cite{BergesonPRA2011,StricklerPRX2016,MelzerPRE2000}.
However, the YOCP cannot be used to describe processes that involve electrons, e.g., ambipolar diffusion or electron-ion temperature relaxation.

The other class of approaches extends the theory of equilibrium density correlations to the case of a plasma with two distinct temperatures.
In an early investigation, Salpeter derived pair correlation functions for a weakly coupled electron-ion plasma using arguments in the vein of Debye-H\"uckel theory~\cite{SalpeterJGR1963}.
Boercker and More extended Salpeter's results to strong ion coupling using an ansatz for a two-temperature partition function, but still required that the electron-ion Coulomb coupling be weak~\cite{BoerckerPRA1986}.
The extension to arbitrary coupling involves a generalization of the theory of pair correlations in equilibrium liquids.
The approaches considered here introduce the notion of a ``cross'' temperature $T_{ab}$ that serves as the kinetic energy scale for inter-species correlations.
One must also determine if the OZ equations themselves should be modified.
An attractive feature of such models is that all species are treated on equal footing, in contrast to the YOCP.
This permits the direct calculation of \textit{all} pair correlation functions and further allows for the possibility of studying two-temperature physics when both species are strongly coupled.

Our main goal is to determine the most accurate approximation available to extend equilibrium integral-equation theories of density correlations to two-temperature plasmas.
There seems to be no consensus at present regarding the form of the cross temperatures, $T_{ab}$, or whether it is necessary to modify the Ornstein-Zernike relations.
This work considers three formulations\cite{SeuferlingPRA1989,BredowCPP2013,DharmaWardanaPRE2008}~that have appeared in recent work on two-temperature strongly coupled plasmas\cite{BredowCPP2013,DharmaWardanaPRE2008,SchwartzCPP2007,RosePOP2009,BenedictPRE2017}.
Our main finding is that the model proposed by Seuferling et al.~(``SVT'') in Ref.~\onlinecite{SeuferlingPRA1989} predicts pair distribution functions that agree with MD over a range of coupling strengths similar to what is seen for the usual equilibrium HNC theory.

We restrict our scope to a plasma with two species of classical point charges, labeled $i$ and $e$, with distinct masses and temperatures.
We focus on testing cases where $T_e \ge T_i$ and $m_e \le m_i$, i.e., the lighter ``electrons'' are warmer than the massive ``ions.''
This is the parameter regime of greatest importance in current strongly coupled plasma contexts.
We also take both species to have equal number density ($n_e=n_i=n/2$) and unit charge $Z_i=Z_e=1$, so that the interaction potential for all particles is the repulsive Coulomb potential,
\begin{equation}
  \label{eq:v-coul}
  v_{ab}(r) = \frac{e^2}{r}~,
\end{equation}
and the Coulomb coupling strength of each species is
\begin{equation}
  \label{eq:gamma}
  \Gamma_s = \frac{e^2/a_s}{\kT_s}~,
\end{equation}
where $a_s=(3/4\pi n_s)^{1/3}$ is the mean spacing between particles of species $s$, $T_s$ is their temperature, $e$ is the elementary charge, and $k_B$ is the Boltzmann constant.

Another basic assumption of this work is the existence of a two-temperature steady state, or ``quasi-equilibrium.''
In a plasma, the collisional exchange of energy tends to be most efficient between particles of the same mass and least efficient between particles of very different mass.
It is frequently the case that particles of each species equilibrate among themselves before the system as a whole relaxes to thermal equilibrium.
On timescales longer than the intraspecies thermal relaxation time but shorter than the interspecies thermal relaxation time, it is often accurate to take the velocity distributions to be Maxwellian with temperatures $T_i$ and $T_e$.

Section~\ref{sec:hnc} introduces the three theories and discusses some of their asymptotic limits.
Section~\ref{sec:md} provides details on the MD techniques used to simulate a two-temperature quasi-equilibrium plasma.
Section~\ref{sec:compare} compares the pair distribution functions of the theoretical models with the MD results.
Section~\ref{sec:yocp} uses the SVT model to study when the YOCP model for ion-ion correlations breaks down as the electron coupling strength increases.
Section~\ref{sec:conc} offers some concluding remarks and describes how the present results will be of use to future studies of two-temperature plasmas.

\section{Candidate HNC Extensions}
\label{sec:hnc}

At thermal equilibrium, the Ornstein-Zernike (OZ) relations are~\cite{OrnsteinKNAW1914}
\begin{equation}
  \label{eq:oz-eq}
  \hat h_{ab}(k) = \hat c_{ab}(k) + \sum_{s=i,e} n_s \hat h_{as}(k) \hat c_{sb}(k)~,
\end{equation}
where $\hat h_{ab}(k)$ and $\hat c_{ab}(k)$ are the Fourier transformed total and direct correlation functions, respectively, and $k$ is the wavenumber.
The OZ equations must be solved in conjunction with approximate closure relations.
The hypernetted-chain (HNC) closure is given by~\cite{vanLeeuwenP1959}
\begin{equation}
  \label{eq:hnc-eq}
  g_{ab}(r) = \exp{ \left[ -\frac{v_{ab}(r)}{\kT} + h_{ab}(r) - c_{ab}(r) \right] }~,
\end{equation}
where $g_{ab}(r) = 1 + h_{ab}(r)$ are the radial distribution functions (RDF).
The HNC closure is very accurate when $v_{ab}(r)$ is long-ranged, as is the case for the Coulomb potential.
However, it is reasonable to expect that the bridge functions (which are neglected in HNC) contribute non-negligibly to the RDFs when $\max{(\Gamma_i,\Gamma_e)} \gtrsim 10$, based on knowledge of the OCP RDFs.
A number of proposed improvements model the neglected bridge functions; see for example Refs.~\cite{RosenfeldPRA1979,IyetomiPRA1983,IyetomiPRA1992,KahlPRE1996}.
We revisit the approximate nature of the HNC closure when comparing with MD results in Sec.~\ref{sec:compare}.

To extend the theory of spatial correlations at equilibrium to multi-temperature systems one must address two points: (a) how to characterize the ``cross-temperatures'' $T_{ab}$ that set the kinetic energy scale for inter-species correlations and (b) whether the OZ relations should be modified.
Most investigations to date have extended the HNC-OZ system of equations in one of three ways.
We will refer to them in this work as the SQRT, MASS, and SVT models.

In the SQRT\cite{BredowCPP2013} and MASS\cite{DharmaWardanaPRE2008} models, the OZ relations are taken to be the same as at equilibrium, and the HNC closures are assumed to be
\begin{equation}
  \label{eq:hnc-Tab}
  g_{ab}(r) = \exp{ \left[ -\frac{v_{ab}(r)}{\kT_{ab}} + h_{ab}(r) - c_{ab}(r) \right] }~.
\end{equation}
The models are distiguished by different ansatzes for the cross-temperatures,
\begin{subequations}
  \begin{align}
    \label{eq:T-sqrt}
    & T^{\msqrt}_{ab} = \sqrt{T_a T_b} \\
    \label{eq:T-mass}
    & T^{\mmass}_{ab} = \frac{m_a T_b + m_b T_a}{m_a + m_b} ,
  \end{align}
\end{subequations}
respectively.
A distinguishing feature of the SQRT model is that it is mass-independent.
In the parameter space of this work ($m_i \ge m_e$, $T_i \le T_e$), it follows that $T^\msqrt_{ei} \le T^\mmass_{ei}$.
Consequently, the SQRT model should be expected to result in stronger interspecies correlations.

In the SVT\cite{SeuferlingPRA1989} model, the cross-temperature is the same as in the MASS model from Eq.~\eqref{eq:T-mass}, but the OZ relations are modified to be
\begin{equation}
  \label{eq:svt-oz}
  \hat h_{ab} = \hat c_{ab} + \sum_{s=i,e} n_s \left( \frac{m_{ab}T_{as}}{m_aT_{ab}} \hat c_{as}\hat h_{sb} + \frac{m_{ab}T_{sb}}{m_bT_{ab}} \hat h_{as}\hat c_{sb} \right) ,
\end{equation}
which we will call the SVT-OZ equations~\footnote{The original formulas in Eq.~(38) of Ref.~\onlinecite{SeuferlingPRA1989} contain some typographical errors.}.
Here, $m_{ab}=m_am_b/(m_a+m_b)$ is the reduced mass of an $a,b$ pair.

The SVT model is based on an ansatz for the two- and three-particle phase-space distribution functions,
\begin{subequations}
\begin{equation}
  \label{eq:bbgky-f2}
  F^{(2)}_{ab} = f_a(p_1) f_b(p_2)~ g_{ab}(r_{12})
\end{equation}
\begin{equation}
  \label{eq:bbgky-f3}
  F^{(3)}_{abc} = f_a(p_1) f_b(p_2) f_c(p_3)~g_{abc}(r_{12}, r_{13}, r_{23})
\end{equation}
\end{subequations}
where
\begin{equation}
  \label{eq:maxwellian}
  f_s(p) = \left(2\pi m_s \kT_s\right)^{-\frac{3}{2}} \exp{\left(\frac{-p^2}{2 m_s \kT_s}\right)}
\end{equation}
is the Maxwell-Boltzmann distribution with temperature $T_s$ normalized to unity, and $g_{abc}$ is the triplet distribution function.
The cross-temperature, $T^\mmass_{ab}$ naturally arises after integrating the two-particle BBGKY equation over momenta, which gives an Yvon-Born-Green-like equation for the RDFs $g_{ab}(r_{12})$ in terms of the triplet functions $g_{abc}(r_{12},r_{13},r_{23})$~\cite{SeuferlingPRA1989,SchwartzCPP2007}.
The SVT-OZ relations are derived by assuming both the superposition approximation for the triplet functions, $g_{abc}\approx g_{ab}g_{ac}g_{bc}$, and the HNC approximation from Eq.~\eqref{eq:hnc-Tab} for the direct correlation functions.
Several steps from this point onward are missing from the derivation in Ref.~\onlinecite{SeuferlingPRA1989}.
These steps are written out in full, in Appendix~\ref{sec:svt-deriv}.

In the MASS and SVT models, the interplay between the mass and temperature dependence of $T_{ei}$ is important.
From Eq.~\eqref{eq:T-mass}, one sees that the mass dependence is dominant, causing $T_{ei}$ to rapidly converge to $T_e$ for $m_i \gtrsim 20 m_e$.
From this, one expects the strength of electron-ion correllations in the MASS and SVT models to be similar to that of the electron-electron correlations when the masses are sufficiently different.

The basic screening physics of each model can be understood through the weakly coupled limit.
In this limit, $\hat c_{ab} \approx -\hat v_{ab}/\kT_{ab}$, and the OZ (or SVT-OZ) equations can be explicitly solved for the partial static structure factors,
\begin{equation}
  \label{eq:sk-def}
  S_{ab}(k) = \delta_{ab} + \sqrt{n_an_b}\,\hat h_{ab}(k)~,
\end{equation}
where $\delta_{ab}$ is the Kronecker delta.
The expressions for each model are written in Appendix~\ref{sec:wc-lim}, from which one can compare the models in both $k$-space and in real space.

First, each model shows qualitative differences in the long wavelength ($k \to 0$) limit.
The values of each model's $S_{ab}(0)$ are tabulated in Table~\ref{tab:Sklim}.
In the long-wavelength limit the SQRT and SVT model structure factors take finite values, as one would expect of a plasma that exhibits Debye screening.
In fact, when $m_e \ll m_i$, both SQRT and SVT give the $S_{ii}(k)$ of a weakly coupled one-component plasma screened by a background species.
However, in the MASS model, $S_{ab}(0)=0$, characteristic of the OCP~\cite{Baus19801}.
The physical content of these differences is further elucidated by examining the charge density structure factor, $S_{ZZ}(k) = \frac{1}{2} ( S_{ii} + 2S_{ei} + S_{ee} )$.
(Note that at weak coupling $S_{ZZ}$ is the same whether the electrons are positively or negatively charged due to the leading sign dependence in $S_{ei}$.)
In the long-wavelength limit, $S_{ZZ}(k)$ describes variations in the total charge density; the condition that the plasma be quasineutral is equivalent to having $S_{ZZ}(0) = 0$.
On the other hand, the long-wavelength limit of the partial structure factors $S_{ab}(0)$ describe screening.
Of the models studied here, only SVT satisfies $S_{ZZ}(0)=0$ with nonzero $S_{ab}(0)$.
Obviously, the MASS model is quasineutral as well, though $S_{ab}(0)=0$ suggests that it does so not by self-consistent screening, but by not allowing long-wavelength density variations of any kind.
The SQRT model is interesting in that its nonzero $S_{ii}(0)$ implies proper Debye screening of ions by electrons, yet it is not quasineutral (SQRT $S_{ZZ}(0) \ne 0$), suggesting that the effect of the ions on the electrons is not consistently treated.

Second, the pole structure of the static structure factors in each model gives rise to different functional forms for the RDFs. 
In the SQRT model, $S_{ab}(k)$ has a single imaginary pole, which leads to an exponentially screened potential after inverse Fourier transformation.
In contrast, the MASS and SVT structure factors have two imaginary poles, so that the screening comes from the difference of two exponentials:
\begin{equation}
  \label{eq:2-yuk}
  g_{ab}(r) \simeq \exp\left\{-\frac{A_1 e^{-K_1r} + A_2 e^{-K_2r}}{4\pi\sqrt{n_an_b}r}\right\} ~.
\end{equation}
Here $A_1$, $A_2$, $K_1$, and $K_2$ are constant coefficients.
Their values for each model are listed in Table~\ref{tab:wc-coeffs}.
Observe that in the limit where $m_e\ll m_i$, the SVT $g_{ii}(r)$ is that of a weakly coupled YOCP screened by electrons.
The relationship between the SVT model and the screened OCP is expounded upon in Section~\ref{sec:yocp}.

%
%
\begin{table}
  \centering
  \begin{tabular}{cccc}
    \hline\hline
    $S_{ab}(0)$ & SQRT & SVT & SVT, $m_e\ll m_i$
    \\ \hline
    ${i-i}$ & $\kappa_e^2/\kappa^2$ &  $(\kappa^2\kappa_{ei}^2 - \kappa_e^2\kappa_i^2)/\kappa^2\kappa_{ei}^2$ & $\kappa_e^2/\kappa^2$
    \\
    ${e-i}$ & $-\kappa_e\kappa_i/\kappa^2$ & $-\frac{m_iT_i+m_eT_e}{m_iT_e+m_eT_i}\kappa_i^2\kappa_e^2/\kappa^2\kappa_{ei}^2$ & -$\kappa_e^2/\kappa^2$ 
    \\
    ${e-e}$ & $\kappa_i^2/\kappa^2$ & $(\kappa^2\kappa_{ei}^2 - \kappa_e^2\kappa_i^2)/\kappa^2\kappa_{ei}^2$ & $\kappa_e^2/\kappa^2$
    \\ \hline\hline
  \end{tabular}
  \caption{Long-wavelength limits of the static structure factors for the weakly coupled limit of the SQRT and SVT models, as well as the SVT model when the mass difference is large. In the MASS model (not listed), all $S_{ab}(0)=0$. The various inverse screening lengths are defined in Appendix~\ref{sec:wc-lim}.}
  \label{tab:Sklim}
\end{table}
\begin{table*}
  \begin{tabular}{c|ccc|ccc|ccc|ccc}
    \hline\hline
    & \multicolumn{3}{|c|}{SQRT}
    & \multicolumn{3}{|c|}{MASS}
    & \multicolumn{3}{|c|}{SVT}
    & \multicolumn{3}{|c}{SVT, $m_e\ll m_i$}
    \\

    & ${i-i}$
    & ${e-i}$
    & ${e-e}$
    & ${i-i}$
    & ${e-i}$
    & ${e-e}$
    & ${i-i}$
    & ${e-i}$
    & ${e-e}$
    & ${i-i}$
    & ${e-i}$
    & ${e-e}$
    \\ \hline
    $A_1$
    & $\kappa_i^2$
    & $\kappa_i\kappa_e$
    & $\kappa_e^2$
    & $\tallfrac{\kappa_+^2(\kappa_i^2 - \kappa_-^2)}{\kappa_+^2-\kappa_-^2}$
    & $\tallfrac{\kappa_{ei}^2\kappa_+^2}{\kappa_+^2-\kappa_-^2}$
    & $\tallfrac{\kappa_+^2(\kappa_e^2 - \kappa_-^2)}{\kappa_+^2-\kappa_-^2}$
    & $\tallfrac{\kappa_i^4}{\kappa^2-\kappa_{ei}^2}$
    & $\tallfrac{\kappa_{ei}^2\kappa^2 
      - c 
      \kappa_i^2\kappa_e^2}{\kappa^2-\kappa_{ei}^2}$
    & $\tallfrac{\kappa_e^4}{\kappa^2-\kappa_{ei}^2}$
    & $\kappa_i^2$
    & $\kappa_{e}^2$
    & $\tallfrac{\kappa_e^4}{\kappa_i^2}$
    \\
    $K_1$
    & $\kappa$
    & $\kappa$
    & $\kappa$
    & $\kappa_+$
    & $\kappa_+$
    & $\kappa_+$ 
    & $\kappa$
    & $\kappa$
    & $\kappa$
    & $\kappa$
    & $\kappa$
    & $\kappa$
    \\
    $A_2$
    & $0$
    & $0$
    & $0$
    & $\tallfrac{\kappa_-^2(\kappa_i^2 - \kappa_+^2)}{\kappa_+^2-\kappa_-^2}$
    & $\tallfrac{\kappa_{ei}^2\kappa_-^2}{\kappa_+^2-\kappa_-^2}$
    & $\tallfrac{\kappa_-^2(\kappa_e^2 - \kappa_+^2)}{\kappa_+^2-\kappa_-^2}$
    & $\tallfrac{\kappa_i^2(\kappa_{ei}^2-\kappa_e^2)}{\kappa^2-\kappa_{ei}^2}$
    & $\tallfrac{\kappa_{ei}^4 
      - c                   
      \kappa_i^2\kappa_e^2}{\kappa^2-\kappa_{ei}^2}$
    & $\tallfrac{\kappa_e^2(\kappa_{ei}^2-\kappa_i^2)}{\kappa^2-\kappa_{ei}^2}$
    & $0$
    & $0$
    & $\tallfrac{\kappa_e^2(\kappa_e^2-\kappa_i^2)}{\kappa_i^2}$
    \\
    $K_2$
    & $-$
    & $-$
    & $-$
    & $\kappa_-$
    & $\kappa_-$
    & $\kappa_-$
    & $\kappa_{ei}$
    & $\kappa_{ei}$
    & $\kappa_{ei}$
    & $-$
    & $-$
    & $\kappa_e$
    \\
    \hline\hline
  \end{tabular}
  \caption{Coefficients appearing in Eq.~\eqref{eq:2-yuk} for the weak-coupling form for the RDFs for each model, as well as for the SVT model when $m_e \ll m_i$. The various inverse screening lengths are defined in Appendix~\ref{sec:wc-lim}, and $c=\frac{m_eT_e+m_iT_i}{m_iT_e+m_eT_i}$ in the SVT column.}
  \label{tab:wc-coeffs}
\end{table*}
%

\section{Simulation Model}
\label{sec:md}

Classical molecular dynamics simulations were carried out using the open-source code LAMMPS\cite{Plimpton1995}.
A two-component, two-temperature plasma was created in a three-dimensional periodic box.
The charged particles were made to interact through the repulsive Coulomb potential, Eq.~\eqref{eq:v-coul}, and the long-range part of the Coulomb interaction was accounted for using the particle-particle, particle-mesh method~\cite{Plimpton97}.

Every simulation system consisted of $10^4$ particles of each species, each singly charged.
The time step for numerical integration was chosen based on the inverse electron plasma frequency, $\omega_{pe}^{-1} = \sqrt{m_e/4\pi e^2 n_e}$.
All simulations used time steps in the range $\delta t =0.005-0.01\omega_{pe}^{-1}$, which was sufficient to resolve the dynamics of both species.

\begin{figure}[t]
  \includegraphics[width=0.9\columnwidth]{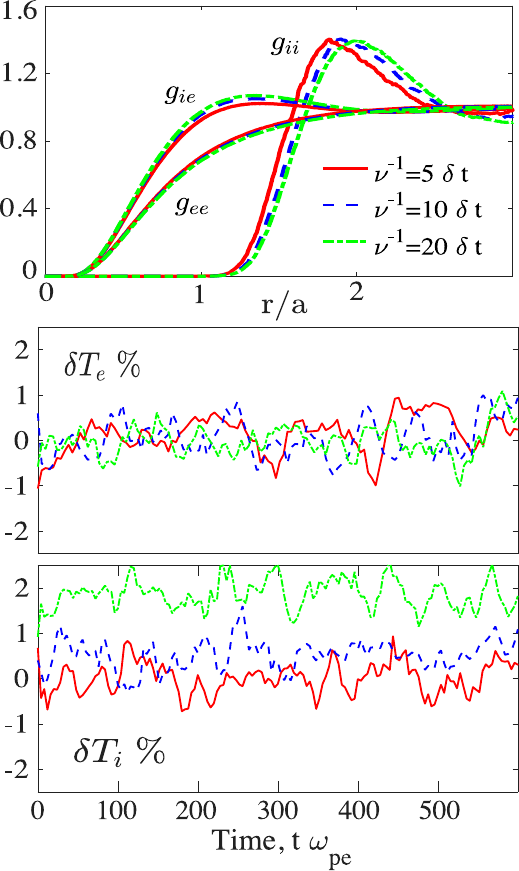}
  \caption{Effects of varying the thermostat Langevin collision frequency on the RDFs and temperature fluctuations. For the simulations shown, $\Gamma_i=50$, $\Gamma_e=1$, and $m_i=30m_e$. Lines show different values of the inverse Langevin collision frequency: $\nu^{-1}=5\delta t$ (solid red), $10\delta t$ (dashed blue), and $20 \delta t$ (dash-dotted green).}
  \label{fig:thermo-coeff}
\end{figure}

The equilibration of a two-species system to two different temperatures remains a nontrivial issue from a numerical point of view~\cite{fukushi_2000}.
For the present simulations, each species was coupled to its own Langevin thermostat.
The Langevin collision frequencies were chosen such that both species attained their target temperatures within $1\%$ statistical fluctuations.
Figure~\ref{fig:thermo-coeff} shows how if the thermostat collision frequency was too weak, the ions thermalized to a temperature that was higher than the target temperature.
One can see that even a $1\%$ drift from the requested $T_i$ is large enough to make a discernible difference in the ion-ion RDF.
We attribute this effect to the fact that in a two-temperature simulation, the thermostats must work against the plasma's natural inclination to thermally relax, which requires that the thermostat collision frequency be greater than the electron-ion collision frequency.
If these two rates are comparable, however, then one expects the ions (which couple to the thermostat inefficiently when their mass is large) to thermalize to a temperature greater than the thermostat temperature but less than the temperature they would attain if allowed to relax.

It was also observed that for high mass ratios, the ion-ion RDF takes much longer to stabilize than the ion temperature.
Even after the ions acquire the temperature of their heat bath, $T_i$, spatial correlations between ions continue to develop for hundreds to thousands of $\omega_{pe}^{-1}$ of simulation time.
In comparison, $g_{ee}$ and $g_{ei}$ stabilize on the same timescale as $T_i$, though small variations thereafter occur in response to the evolution of $g_{ii}$.
For reference, the OCP typically requires only a few plasma periods of averaging time for well-resolved RDFs.
Since the case of large mass ratio is of particular experimental importance, the computational burden of simulating such plasmas underscores the need for a reliable theoretical model of correlations in two-temperature plasmas.

We compared the RDFs obtained from a system under Langevin thermostats with those of a system equilibrated using two simultaneous Nos\'{e}-Hoover thermostats.
At higher mass ratios, the results remain identical irrespective of choice of thermostat.
At lower mass ratios ($m_i/m_e \lesssim 5$), the system under Nos\'{e}-Hoover thermostats displayed the ``flying ice cube effect,'' in which the system accumulated a spurious net momentum, leading to incorrect RDFs~\cite{harvey_98}.
The Langevin thermostats, however, were found to give consistent RDFs for all mass ratios.

The simulations were carried out in three stages.
First, we performed an initial thermostatting stage until each species reached its target temperature.
The required length of this phase depended on the mass ratio.
It was found that for $m_i=m_e$, $400$ electron plasma periods were sufficient and that this number scaled with increased ion mass as $\sqrt{m_i/m_e}$.
Second, the evolution of the RDFs was monitored until it was seen that the ion-ion correlations had fully developed.
Third, time-averaged RDFs were computed while keeping both the thermostats on.
The thermostats were kept active to prevent electron-ion temperature relaxation over the timescales necessary to accurately sample the RDFs.
Because the thermostats were left on during the entire simulation period, the total energy was not conserved.

\section{Comparison of HNC with MD}
\label{sec:compare}

\begin{figure*}
  \includegraphics[width=\textwidth]{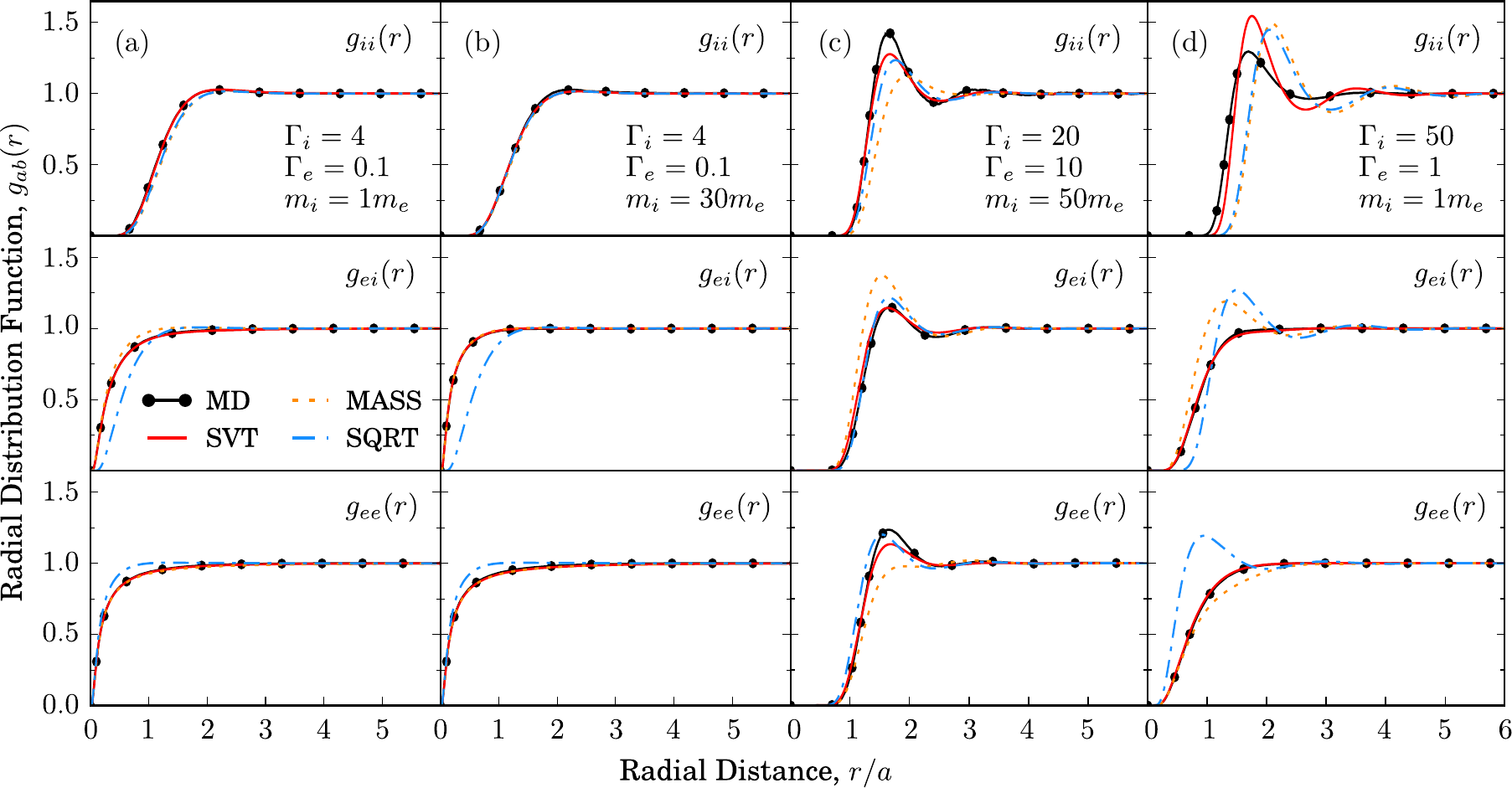}
  \caption{Model RDFs compared with molecular dynamics simulation results. Connected black circles are MD, solid red lines are the SVT model, dotted orange lines are the MASS model, and dash-dotted blue lines are the SQRT model.}
  \label{fig:rdfs}
\end{figure*}

We have evaluated each of the three HNC extensions described in Section~\ref{sec:hnc} and conducted MD simulations as described in Section~\ref{sec:md} for several combinations of coupling strengths and mass ratios.
Here we present an illustrative subset of the comparisons made, shown in Figure~\ref{fig:rdfs}.
Plots for other parameter combinations can be found the Supplementary Material.

Figures~\ref{fig:rdfs}a-b show the radial distribution functions for a plasma of strongly coupled ions and weakly coupled electrons, with $m_i=m_e$ and $m_i=30m_e$.
The first observation to make is that the strength of electron-ion correlations is clearly set by $T_{ei}^\mmass$, not by $T_{ei}^\msqrt$.
The too-wide Coulomb hole in $g_{ei}(r)$ shows that the SQRT model overestimates the strength of electron-ion coupling.
Furthermore, the SQRT model predicts electron-electron correlation functions that qualitatively differ from MASS, SVT, and MD.
The physical reason is most clearly illustrated by examining the weakly coupled limit of the SVT $g_{ee}(r)$ when $m_e \ll m_i$.
Identifying the potential of mean force as $\phi_{ee} = -\kT_e\ln g_{ee}$, one can write (see Eq.~\eqref{eq:2-yuk} and Table~\ref{tab:wc-coeffs})
\begin{equation}
  \label{eq:svt-masslim-gee}
  \phi^\msvt_{ee}(r) \simeq \frac{e^2}{r} 
  \left[ e^{-\kappa_e r} - \frac{\kappa_e^2}{\kappa_i^2}
    \left(e^{-\kappa_e r} - e^{-\kappa r}
    \right) 
  \right]~.
\end{equation}
The first term is the screened repulsion that electrons would experience from one another if they were an OCP, while the ``attractive'' second term results from the tendency for electrons to cluster when they form screening clouds around ions.
These two processes compete, giving rise to the slow decay in the SVT, MASS, and MD $g_{ee}(r)$ compared to the SQRT model, which lacks this second ``attractive'' part.
These deficiencies in the SQRT $g_{ei}(r)$ and $g_{ee}(r)$ were present at all coupling strengths and mass ratios investigated.
The errors between SQRT and MD worsen at stronger coupling strengths, as can be seen in the Supplementary Material.

\begin{figure}
  \includegraphics[width=0.9\columnwidth]{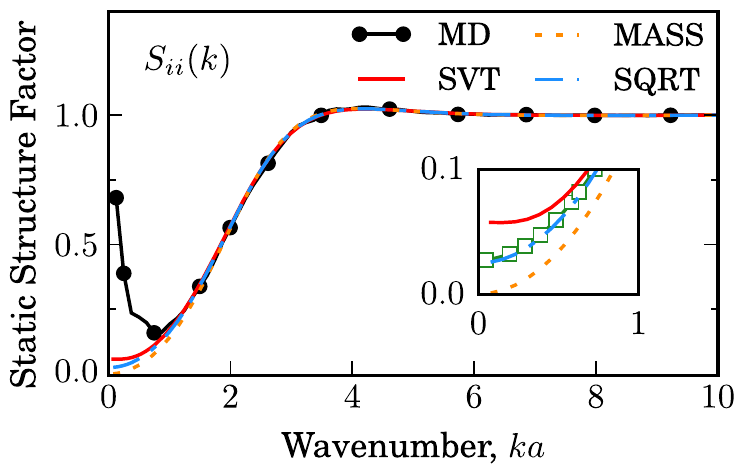}
  \caption{Model ion-ion static structure factors compared with molecular dynamics simulation for $\Gamma_i=4$, $\Gamma_e=0.1$, and $m_i=30m_e$. The inset shows $S_{ii}(k)$ near $k=0$, including the YOCP model (green squares).}
  \label{fig:sk-gmi-4-gme-0p1-mime-30}
\end{figure}

The remaining comparison of the MASS and SVT models highlights the question of whether the OZ equations require modification to describe a two-temperature system.
In all cases studied, the SVT radial distribution functions more closely agree with MD, though the differences between the MASS and SVT RDFs often appear small.
In fact, in Ref.~\onlinecite{DharmaWardanaPRE2008}, the MASS model's apparent accuracy is cited as evidence that SVT's modified OZ equations are unnecessary.
Important differences in favor of the SVT approach surface when comparing the structure factors.
An example is shown in Figure~\ref{fig:sk-gmi-4-gme-0p1-mime-30}.
The ion-ion structure factor vanishes in the MASS model as $k\to 0$, indicating that the ions are thermodynamically similar to an \textit{unscreened} OCP, despite the presence of a screening electron background.
In contrast, the SVT model gives a finite value, in line with both MD and the YOCP model.
This behavior is demonstrated analytically in Sec.~\ref{sec:yocp}.

Since all the models considered are variants of the HNC approximation, it should be expected that they will all suffer inaccuracies at higher coupling due to the lack of bridge functions.
In the OCP, bridge functions primarily correct the RDF oscillation amplitudes, which are somewhat too small without the bridge functions.
Other differences such as the size of the Coulomb hole and the oscillation phase are relatively minor, so if these features are a point of disagreement between the models and MD, it is more likely due to the two-temperature modeling than the lack of bridge functions.

Figure~\ref{fig:rdfs}c shows the RDFs when both species are strongly coupled.
As expected, the SVT model underestimates the peak of $g_{ii}(r)$ but otherwise agrees well with MD.
In contrast, the MASS model appears to break down entirely in this regime of strong electron coupling.
An unexpected feature of the MD RDFs is that at high mass ratio, the height of the first peak of $g_{ee}$ exceeds that of $g_{ei}$.
Ordinarily, one expects the height of this peak to correlate with the strength of the bare interaction compared to the kinetic energy, so that since $T_i < T_{ei} < T_e$, one anticipates $\max{(g_{ii})} > \max{(g_{ei})} > \max{(g_{ee})}$.
For low mass ratios, both MD and the HNC models bear out this trend at all coupling strengths, while at higher mass ratios, the HNC models do not capture the augmented first correlation peak in $g_{ee}$ observed in MD.

Figure~\ref{fig:rdfs}d shows the breakdown of the two-temperature HNC models at higher ion coupling strength.
All three two-temperature models \textit{over}estimate the strength of correlations in the plasma, exhibiting Coulomb holes and RDF oscillations that are larger than those seen in the MD simulations.
This is in contrast to the usual equilibrium HNC theory fails, which underpredicts the peaks.
For higher mass ratios and/or lower temperature ratios (see the Supplementary Material), the SVT RDFs are in surprisingly good agreement with MD even at such strong coupling.
These are cases that happen to lie in the transitional regime where SVT goes from underpredicting to overpredicting the RDF peaks.

%

\section{Comparison with the Yukawa OCP }
\label{sec:yocp}

We now compare the ion-ion correlations of the SVT model to the Yukawa OCP to test the YOCP's limitations as $\Gamma_e$ increases.
In the classical YOCP model, the electrons are an ideal background that screens the ions.
The ions then interact through a Debye-screened potential,
\begin{equation}
  \label{eq:v-yuk}
  v^\y_{ii}(r) = \frac{e^2}{r} e^{-\kappa_e r}~,
\end{equation}
where $\kappa_e=\sqrt{3\Gamma_e}a_e^{-1}$ is the inverse electron Debye length.
The YOCP model is valid only when the screening background is weakly coupled, while the SVT model predicts accurate ion-ion RDFs even when $\Gamma_e$ exceeds unity.
By comparing the YOCP ion-ion RDF $g_\y(r)$ with $g_{ii}(r)$ from two-temperature SVT calculations, we can quantitatively assess at what $\Gamma_e$ the YOCP model fails.

A result of the weak-coupling approximation is that $\kappa_e$ does not depend on the sign of the electron charge.
For this reason, the weak-coupling assumption of the YOCP can be tested using positively charged electrons in the SVT calculations; however, an important caveat must be made.
As the electron coupling strength increases, the nature of how they screen the ions is expected to become increasingly dependent on the sign of their charge.
It is reasonable to expect, though, that the $\Gamma_e$ at which the exponential screening approximation fails is about the same value at which the sign of the electron charge becomes important, since they are both tied to the weak-coupling assumption.
We expect, then, that the $\Gamma_e$ threshold reported here should not strongly depend on the use of positively charged electrons.

\begin{figure}
  \includegraphics[width=\columnwidth]{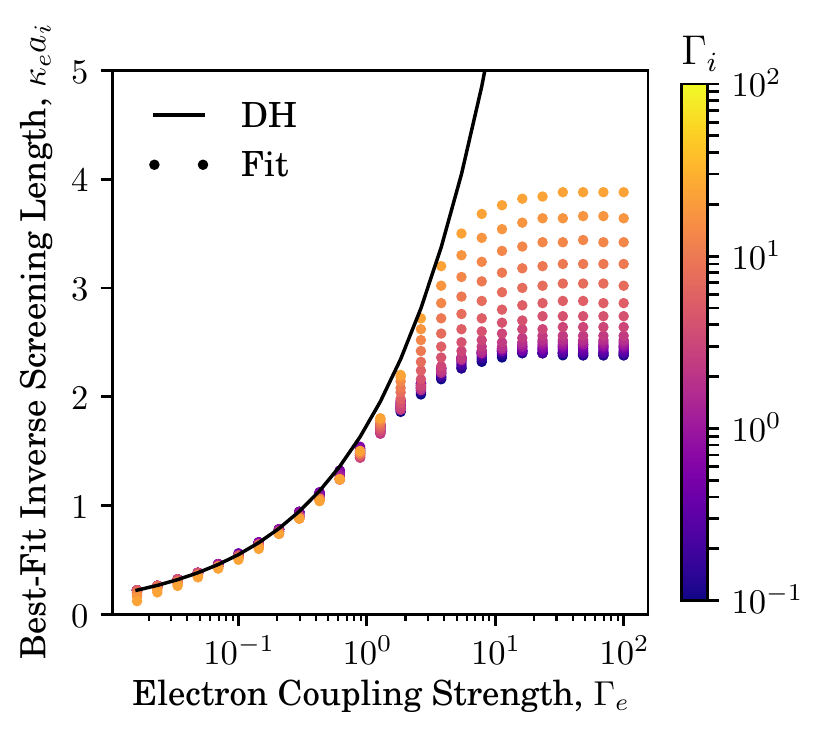}
  \caption{Inverse screening length of the YOCP whose $g_\y(r)$ (pink) best matches the SVT $g_{ii}(r)$ at the same ion coupling strength. The fit criterion is given by Eq.~\eqref{eq:fit}. Multiple points at the same $\Gamma_e$ are for different values of $\Gamma_i$.}
  \label{fig:yocp-fit}
\end{figure}

For a given $\Gamma_i$, we solve the HNC-SVT-OZ equations for $g_{ii}(r)$ at several $\Gamma_e$ and solve the ordinary HNC-OZ equations for $g_\y(r)$ at several $\kappa_e$.
For each $\Gamma_e$, the best-fit $\kappa_e$ was chosen to be the one that minimizes the integrated absolute difference between $g_\y$ and $g_{ii}$ from HNC,
\begin{equation}
  \label{eq:fit}
  \Delta = \int d\vec r |g_\y(r;\kappa_e) - g_{ii}(r)|~.
\end{equation}
%
Figure~\ref{fig:yocp-fit} shows the best-fit YOCP $\kappa_e$ over a wide range in $\Gamma_i$ and $\Gamma_e$ with the mass ratio fixed at $m_i=1836m_e$.
Immediately, one sees that when the electrons are weakly coupled, the best-fit $\kappa_e$ is independent of the ion coupling strength and furthermore is essentially the inverse electron Debye length, plotted in black in the figure.
The reason becomes clear upon investigating the SVT-OZ equations at weak electron coupling.

In the limit of weak electron coupling, the Debye-H\"uckel approximation should be excellent for the electron-electron direct correlation function.
Since $T_{ei}\approx T_e$, the same should be true of the electron-ion direct correlation function, giving
\begin{equation}
  \label{eq:c-yocp}
  \hat c_{ee}(k) \approx Z_i^{-1}\hat c_{ei}(k) \approx -\frac{4\pi e^2}{\kT_e}\frac{1}{k^2}~.
\end{equation}
Due to the large mass ratio, the SVT-OZ equations from Eq.~\eqref{eq:svt-oz} become
\begin{subequations}
  \label{eq:svt-masslim}
  \begin{align}
    & \hat h_{ii} = \hat c_{ii} + n_i \hat h_{ii} \hat c_{ii} + n_e \frac{T_e}{T_i} \hat h_{ei} \hat c_{ei}
    \\
    & \hat h_{ei} = \hat c_{ei} + n_i \hat h_{ii} \hat c_{ei} + n_e \hat h_{ei} \hat c_{ee}
    \\
    & \hat h_{ee} = \hat c_{ee} + n_i \hat h_{ei} \hat c_{ei} + n_e \hat h_{ee} \hat c_{ee}~.
  \end{align}
\end{subequations}
Since $\hat c_{ei}$ and $\hat c_{ee}$ are known, $h_{ei}$ can be eliminated from the first equation to find
\begin{equation}
  \label{eq:oz-scr}
  \hat h_{ii} = 
  \left(
    \hat c_{ii} + \frac{n_e\hat c_{ee}}{1 - n_e\hat c_{ee}} \frac{T_e}{T_i} \hat c_{ei}
  \right)
  \left(
    1 + n_i \hat h_{ii}
  \right)~.
\end{equation}
If we introduce the notion of the ``screened'' ion-ion direct correlation function
\begin{equation}
  \label{eq:c-scr}
  \hat c_{\scr}
  = \hat c_{ii} + \frac{n_e\hat c_{ee}}{1 - n_e\hat c_{ee}} \frac{T_e}{T_i} \hat c_{ei}~,
\end{equation}
then the ion structure factor is given by
\begin{equation}
  \label{eq:svt-dh-sii}
  S_{ii}(k) = \frac{1}{1 - n_i\hat c_\scr(k)}~,
\end{equation}
meaning that $\hat c_\scr$ mediates a one-to-one mapping between the ion structure of the two-component plasma and that of an equivalent screened one-component plasma.

In the Debye-H\"uckel approximation for the electrons,
\begin{equation}
  \label{eq:c-scr-dh}
  n_i \hat c_\scr = n_i \hat c_{ii} + \frac{T_i}{T_e} \frac{1}{\lambda_{Di}^2k^2}\frac{1}{1 + \lambda_{De}^2k^2}~.
\end{equation}
Now if we decompose $\hat c_{ii}(k)$ into its singular Coulombic part and a remainder $\hat c^{R}_{ii}=\hat c_{ii} + \hat v_{ii}/\kT_i$ that is regular as $k \to 0$~\cite{Baus19801,BausJPA1978}, we find 
\begin{equation}
  n_i \hat c_\scr(k) = n_i\hat c_{ii}^R(k) - \frac{\lambda_{De}^2}{\lambda_{Di}^2} \frac{1}{1 + \lambda_{De}^2k^2}~.
\end{equation}
Thus the long wavelength limit of the ion structure factor is
\begin{equation}
  \lim_{k\to 0} S_{ii}(k) = \frac{1}{1 - n_i\hat c_{ii}^{R}(0) + (\lambda_{De}/\lambda_{Di})^2}~.
\end{equation}
Repeating these steps using the ordinary OZ equations results in Eq.~\eqref{eq:c-scr-dh}, but without the factor of $T_i/T_e$, which causes $\hat c_\scr$ to remain singular.
This is the reason why the MASS model structure factor is zero in the $k \to 0$ limit, while the same limit in the SVT model is YOCP-like (nonzero).
In passing, it is interesting to note that inserting Eq.~\eqref{eq:svt-dh-sii} and~\eqref{eq:c-scr-dh} back into Eq.~\eqref{eq:svt-masslim} give the same structure factors found by Boercker and More~\cite{BoerckerPRA1986}.

\begin{figure}
  \includegraphics[width=\columnwidth]{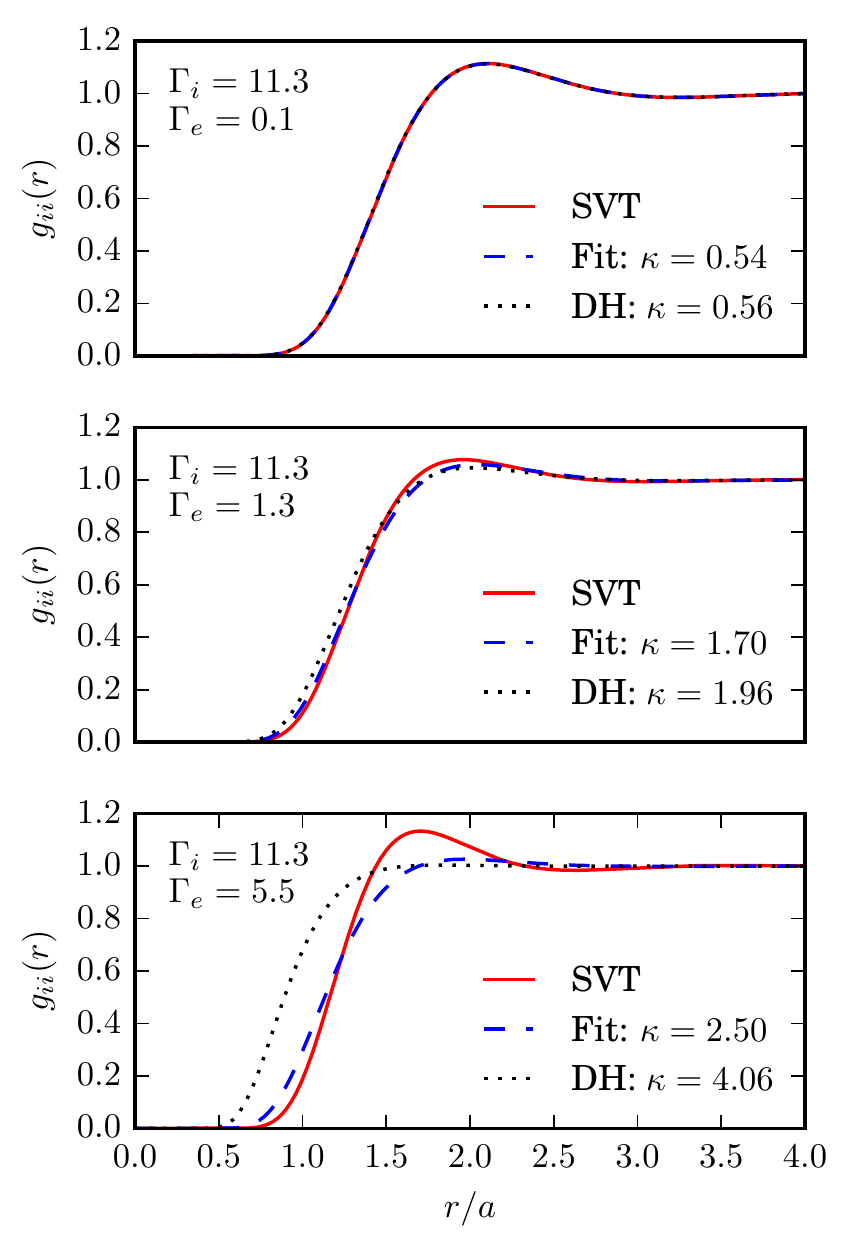}
  \caption{Comparison of ion-ion RDFs obtained from SVT (solid red), YOCP with fitted $\kappa_e$ (dashed blue), and YOCP with $\kappa_e$ equal to the inverse electron Debye length (dotted black).}
  \label{fig:yocp-fit-gr-cmp}
\end{figure}

Figure~\ref{fig:yocp-fit-gr-cmp} demonstrates the breakdown of YOCP behavior when the electrons become strongly coupled.
Interestingly, even when $\Gamma_e \simeq 1$, both the fitted YOCP model and the Debye-H\"uckel model are in fair agreement with the full two-component SVT calculation.
However, further increases to $\Gamma_e$ result in ion-ion RDFs that rapidly become non-YOCP-like; even the fitted YOCP underpredicts the ion-ion correlation strength.
In other words, the mapping between the two-component system and effective one-component system given by Eq.~\eqref{eq:c-scr} can no longer be reproduced by an effective Yukawa potential.

%
%

\section{Conclusions }
\label{sec:conc}

By comparison with molecular dynamics simulations, it has been demonstrated that the model proposed by Seuferling, Vogel, and Toeppfer~\cite{SeuferlingPRA1989} accurately extends the Ornstein-Zernike theory of pair correlations to two-temperature plasmas up to and slightly beyond the coupling strengths achieved by present-day ultracold neutral plasma experiments.
The assumption of a mass-weighted ``cross-temperature'' correctly predicts the suppression of electron-ion correlations when the mass ratio is large.
Further, we have shown that the modifications made by SVT to the Ornstein-Zernike equations are necessary to give nonzero long-wavelength limits of the static structure factors, which correctly reflects the self-consistent screening of ions by electrons and vice-versa.
These findings are given additional weight by our direct comparisons of the ion-ion correlation functions in the SVT and Yukawa OCP models, which indicate that the Yukawa OCP model will become unsuitable even for modeling ion correlations once $\Gamma_e\gtrsim 1$.

The present work marks important progress towards a fully two-component description of correlations in classical strongly coupled plasmas.
In particular, it suggests that the SVT model can be used to obtain accurate effective scattering potentials or static local field corrections needed in quasi-static descriptions of transport and relaxation processes of strongly coupled plasmas\cite{BaalrudPRL2013,IchimaruPRA1985,DharmaWardanaPRE1998,DaligaultPRE2009,VorbergerPRE2010,BenedictPRE2017}.
However there remain interesting physical challenges to overcome.
Future work will address the issue of the electron charge, which was taken to be positive in this work to decouple the relevant two-temperature physics from the physics of classical recombination.
There is also the question of how to best simulate a two-temperature steady state, both in terms of technical choices regarding thermostats and in terms of the basic statistical mechanics of the simulated ensemble.

\section*{Supplementary Material}

See supplementary material for plots of the radial distribution functions and static structure factors for all coupling strengths and mass ratios investigated in this work.

\begin{acknowledgments}

This material is based upon work supported by the National Science Foundation under Grant No.~PHY-1453736 and by the Air Force Office of Scientific Research under award number FA9550-16-1-0221.
It used the Extreme Science and Engineering Discovery Environment (XSEDE), which is supported by NSF Grant No.~ACI-1053575, under Project Award No.~PHY-150018.


\end{acknowledgments}

\appendix{}



\section{Derivation of the SVT-OZ Equations}
\label{sec:svt-deriv}

We begin from Eq.~(5) of Ref.~\onlinecite{SchwartzCPP2007}, which after applying the superposition approximation, $g_{abc}\approx g_{ab}g_{ac}g_{bc}$, can be written
\begin{equation}
  \label{eq:ybg}
  \begin{split}
    &
    \pp{}{\vec r_{1}} \big[ \kT_{ab}\ln{g_{ab}} + v_{ab} \big]
    \\
    & \quad
    = - \sum_{c} n_c \int d\vec r_3
    \left[
      \frac{m_{ab}}{m_a} \pp{v_{ac}}{\vec r_1}
      -
      \frac{m_{ab}}{m_b} \pp{v_{bc}}{\vec r_2}
    \right]
    g_{ac} g_{bc} ~,
\end{split}
\end{equation}
where particles $1$ and $2$ are of species $a$ and $b$, respectively (which could be the same or different), and the sum runs over all species labels.
Eq.~\eqref{eq:ybg} is a closed set of equations for the RDFs, but it is not suitable for strongly coupled systems because of the use of the superposition approximation.
One introduces the direct correlation functions through an HNC-like approximation,
\begin{equation}
  \ln g_{ab} = -\frac{v_{ab}}{\kT_{ab}} + h_{ab} - c_{ab}~,
\end{equation}
in the hope that the errors from HNC will cancel somewhat the errors made by the superposition approximation.
With this and the fact that the lack of external forces implies
\begin{equation}
  \sum_c n_c \int d\vec r_3 g_{ac}\pp{v_{ac}}{\vec r_1} = \sum_c n_c \int d\vec r_3 g_{bc}\pp{v_{bc}}{\vec r_2} = 0
\end{equation}
Eq.~\eqref{eq:ybg} becomes
\begin{equation}
  \begin{split}
    & \pp{}{\vec r_1} \big[ h_{ab}(\vec r_{12}) - c_{ab}(\vec r_{12}) \big]
    \\
    & \quad =
    \sum_c n_c \frac{m_{ab}}{m_a}\frac{T_{ac}}{T_{ab}} \pp{}{\vec r_1} [c_{ac} \star h_{bc}](\vec r_{12})
    \\
    & \quad +
    \sum_c n_c \frac{m_{ab}}{m_b}\frac{T_{bc}}{T_{ab}} \pp{}{\vec r_1} [h_{ac} \star c_{bc}](\vec r_{12})
    \\
    & \quad -
    \sum_c n_c \frac{m_{ab}}{m_a}\frac{T_{ac}}{T_{ab}} \int d\vec r_3 h_{ac}(\vec r_{13})\pp{\gamma_{ac}(\vec r_{13})}{\vec r_{1}}h_{bc}(\vec r_{23})
    \\
    & \quad +
    \sum_c n_c \frac{m_{ab}}{m_b}\frac{T_{bc}}{T_{ab}} \int d\vec r_3 h_{bc}(\vec r_{23})\pp{\gamma_{bc}(\vec r_{23})}{\vec r_{2}}h_{ac}(\vec r_{13})~,
\end{split}
\end{equation}
where $\gamma = h - c$ is the indirect correlation function, and the $\star$ operation denotes convolution.
After Fourier transforming $\vec r_1\to\vec k,\vec r_2\to\vec k'$, integrating over $\vec k'$, and dotting $\vec k$ on both sides, one obtains
\begin{equation}
  \begin{split}
    &
    k^2
    \left[
      \hat h_{ab}(\vec k) - \hat c_{ab}(\vec k)
    \right]
    \\
    & \quad =
    k^2 \sum_c n_c \frac{m_{ab}}{m_a}\frac{T_{ac}}{T_{ab}} \hat c_{ac}(\vec k) \hat h_{bc}(-\vec k)
    \\
    & \quad +
    k^2 \sum_c n_c \frac{m_{ab}}{m_b}\frac{T_{bc}}{T_{ab}} \hat h_{ac}(\vec k) \hat c_{bc}(-\vec k)
    \\
    & \quad -
    \sum_c n_c \frac{m_{ab}}{m_a}\frac{T_{ac}}{T_{ab}} \hat h_{bc}(-\vec k)
    \int d\vec \ell (\vec k \cdot \vec \ell) \hat \gamma_{ac}(\vec \ell)\hat h_{ac}(\vec k - \vec  \ell)
    \\
    & \quad +
    \sum_c n_c \frac{m_{ab}}{m_b}\frac{T_{bc}}{T_{ab}} \hat h_{ac}(\vec k)
    \int d\vec \ell (\vec k \cdot \vec \ell) \hat \gamma_{bc}(\vec \ell)\hat h_{bc}(-\vec k - \vec \ell)~,
\end{split}
\end{equation}
where $\vec \ell$ is a dummy wavenumber arising from the Fourier transform of a real-space product.
Since all the correlation functions must be isotropic in their arguments,
\begin{equation}
  \label{eq:svt-plus-extra}
  \begin{split}
    &
    k^2
    \left[
      \hat h_{ab}(k) - \hat c_{ab}(k)
    \right]
    \\
    & \quad =
    k^2 \sum_c n_c \frac{m_{ab}}{m_a}\frac{T_{ac}}{T_{ab}} \hat c_{ac}(k) \hat h_{bc}(k)
    \\
    & \quad +
    k^2 \sum_c n_c \frac{m_{ab}}{m_b}\frac{T_{bc}}{T_{ab}} \hat h_{ac}(k) \hat c_{bc}(k)
    \\
    & \quad -
    \sum_c n_c \frac{m_{ab}}{m_a}\frac{T_{ac}}{T_{ab}} \hat h_{bc}(k)
    \int d\vec \ell (\vec k \cdot \vec \ell) \hat \gamma_{ac}(\ell)\hat h_{ac}(|\vec k - \vec\ell|)
    \\
    & \quad +
    \sum_c n_c \frac{m_{ab}}{m_b}\frac{T_{bc}}{T_{ab}} \hat h_{ac}(k)
    \int d\vec \ell (\vec k \cdot \vec \ell) \hat \gamma_{bc}(\ell)\hat h_{bc}(|\vec k - \vec \ell|)~,
\end{split}
\end{equation}
where we have taken $\vec \ell \to -\vec \ell$ in the last line.
The first three lines together form $k^2$ times the SVT-OZ equations as written in Eq.~\eqref{eq:svt-oz}, so the remaining two terms must vanish.
We abbreviate
\begin{equation}
  z_{ab}(k) = k^{-2}\int d\vec \ell (\vec k \cdot \vec \ell)\hat \gamma_{ab}(\ell)\hat \gamma_{ab}(|\vec k - \vec \ell|)~,
\end{equation}
and call the last two terms of Eq.~\eqref{eq:svt-plus-extra} the ``remainder,'' $R_{ab}$, so that Eq.~\eqref{eq:svt-plus-extra} may be written
\begin{equation}
  \label{eq:svt-plus-rem}
  \begin{split}
    \hat h_{ab}
    =
    \hat c_{ab}
    & +
    \sum_c n_c \frac{m_{ab}}{m_a}\frac{T_{ac}}{T_{ab}} \hat c_{ac} \hat h_{bc}
    \\
    & +
    \sum_c n_c \frac{m_{ab}}{m_b}\frac{T_{bc}}{T_{ab}} \hat h_{ac} \hat c_{bc}
    + R_{ab}
  \end{split}
\end{equation}
with
%
  \begin{align}
    \begin{split}
      R_{ab} 
      & =
      - \sum_c n_c \frac{m_{ab}}{m_a}\frac{T_{ac}}{T_{ab}} \hat h_{bc}(k) \hat z_{ac}(k)
      \\
      & \quad
      + \sum_c n_c \frac{m_{ab}}{m_a}\frac{T_{ac}}{T_{ab}} \hat h_{ac}(k) \hat z_{bc}(k)
    \end{split}
  \end{align}
%
For the like-species equation ($a=b$), $R_{ab}$ vanishes trivially.
For the cross-species equation ($a\ne b$), observe that $R_{ab}$ changes sign upon interchange of species labels ($a\leftrightarrow{}b$), while the other terms of Eq.~\eqref{eq:svt-plus-rem} do not.
Therefore, $R_{ab}=0$ for all combinations of $a$ and $b$, giving Eq.~\eqref{eq:svt-oz}.
\section{The Weakly Coupled Limit}
\label{sec:wc-lim}

In the limit of weak coupling, the direct correlation functions may be approximated
\begin{equation}
  \hat c_{ab}(k) \approx -\frac{4\pi e^2}{\kT_{ab}}\frac{1}{k^2}~,
\end{equation}
and it is straightforward to solve each model for the static stucture factors, $S_{ab}(k) = \delta_{ab} + \sqrt{n_an_b}\hat h_{ab}(k)$, in terms of various characteristic screening lengths.
Using the notation
\begin{align*}
  & \kappa^2_i = 4\pi e^2 n_i / \kT_i \\
  & \kappa_e^2 = 4\pi e^2 n_e / \kT_e \\
  & \kappa_{ei}^2 = 4\pi e^2 \sqrt{n_in_e} / \kT^\mmass_{ei} \\
  & \kappa^2 = \kappa_e^2 + \kappa_i^2 \\
  & \kappa_{\pm}^2 = \frac{\kappa^2}{2} \pm \sqrt{\frac{\kappa^4}{4} - \kappa_i^2\kappa_e^2 + \kappa_{ei}^4}~,
\end{align*}
one finds for the SQRT model,
\begin{subequations}
  \label{eq:wc-sqrt}
  \begin{align}
    & S_{ii} = \frac{k^2 + \kappa_e^2}{k^2 + \kappa^2}
    \\
    & S_{ei} = \frac{-\kappa_e\kappa_i}{k^2 + \kappa^2}
    \\
    & S_{ee} = \frac{k^2 + \kappa_i^2}{k^2 + \kappa^2}
  \end{align}
\end{subequations}
for the MASS model,
\begin{subequations}
  \label{eq:wc-mass}
  \begin{align}
      & S_{ii} = \frac{
        k^4 + \kappa_e^2 k^2
        }{
        (\kappa^2 + \kappa_+^2)(\kappa^2 + \kappa_-^2)
        } \\
      & S_{ei} = \frac{
        -\kappa_{ei}^2k^2
        }{
        (\kappa^2 + \kappa_+^2)(\kappa^2 + \kappa_-^2)
        } \\
      & S_{ee} = \frac{
        k^4 + \kappa_i^2 k^2
        }{
        (\kappa^2 + \kappa_+^2)(\kappa^2 + \kappa_-^2)
        }
  \end{align}
\end{subequations}
and for the SVT model,
\begin{subequations}
  \label{eq:wc-svt}
  \begin{align}
      & S_{ii} = \frac{
        (k^2 + \kappa_{ei}^2)(k^2 + \kappa_e^2) + \kappa_i^2(\kappa_{ei}^2 - \kappa_e^2)
        }{
        (k^2 + \kappa^2)(k^2 + \kappa_{ei}^2)
        } \\
      & S_{ei} = \frac{
        -\kappa_{ei}^2k^2 - \frac{m_iT_i+m_eT_e}{m_iTe+m_eT_i}\kappa_e^2\kappa_i^2
        }{
        (k^2 + \kappa^2)(k^2 + \kappa_{ei}^2)
        } \\
      & S_{ee} = \frac{
        (k^2 + \kappa_{ei}^2)(k^2 + \kappa_i^2) + \kappa_e^2(\kappa_{ei}^2 - \kappa_i^2)
        }{
        (k^2 + \kappa^2)(k^2 + \kappa_{ei}^2)
        }~.
  \end{align}
\end{subequations}

\bibliography{refs}

\end{document}